\documentclass[aps,prd,showpacs,twocolumn,amsfonts]{revtex4}
\usepackage{amsmath,pstricks,pst-node,psfig}

\begin{document}

\title{Pion gas viscosity at low temperature and density.}
\author{Antonio Dobado}\email{malcon@fis.ucm.es}
\affiliation{Departamento de F\'{\i}sica Te\'orica,\\
Universidad Complutense de Madrid, 28040 Madrid, Spain}
\author{Silvia N. Santalla}\email{ssantall@fis.uc3m.es}
\affiliation{Departamento de F\'{\i}sica,\\
Universidad Carlos III de Madrid, 28911 Legan\'es, Spain}

\begin{abstract}
By using Chiral Perturbation Theory and the Uehling-Uhlenbeck
equation we compute the viscosity of a pion gas, in the low
temperature and low density regime, in terms of the temperature,
and the pion fugacity. The viscosity turns out to be proportional
to the squared root of the temperature over the pion mass. Next to
leading corrections are proportional to the temperature over the
pion mass to the $3/2$.
\end{abstract}
 \maketitle


\section{Introduction}

The possibility of discovering the so called quark-gluon plasma at
RHIC (Relativistic Heavy Ion Collider) or in the future LHC (Large
Hadron Collider) has triggered a lot of effort in the theoretical
description of ultra-relativistic heavy ion collisions. The modern
view of these collisions was set mainly by Bjorken \cite{Bjorken}
and it is based in the hydrodynamic model already considered by
Landau \cite{Landau1}. In the last stages of the evolution of the
central rapidity region the hadronic fluid is  made mainly of
pions. It has been argued that this expanding pion gas can reach
thermal equilibrium  much more efficiently than chemical
equilibrium since at low energies pion interactions are mostly
elastic \cite{Leutwyler}. In order to have an appropriate
phenomenological description of the expanding pion gas it is
needed to set the proper hydrodynamic equations. In the pioneering
work by Bjorken the hadronic gas (both in the confined and in the
quark-gluon plasma phases) was assumed to be a perfect
relativistic fluid and accordingly viscosity and heat conductivity
were neglected. In order to check the validity of this assumption
in the different regimes of the pion gas expansion it is needed to
compute the corresponding transport coefficients as a function of
the temperature and density. This could make possible to know in
what physical situations they are relevant and the ideal fluid
equations are not appropriate any more. In addition the
expressions obtained for these coefficients can be used to set the
correct relativistic version of the Navier-Stokes equations which
applies when dissipative effects must be taken into account in the
pion fluid dynamics. Curiously  enough, the computation of these
coefficients is also interesting from the fundamental point of
view, since it can be done completely from first principles.

In this work we will illustrate this by computing the viscosity of
the pion gas at low energy and density. Our computation will rely
just in standard kinetic theory and the chiral symmetry properties
of the strong interaction and thus our results can be considered
model independent. In fact our results will depend only on the
pion mass $M_\pi$ and the pion decay constant $F_\pi$. The main
points on which our computation is based are the following. First
of all our computation is going to apply only at low temperature
and density. At low temperatures hadronic matter is confined into
hadrons and at very low temperatures the only excited modes are
pions for vanishing barionic number $n_B$ density. Notice that
$n_B\sim 0$  is supposed to be the case in the central rapidity
region of heavy ion collisions. Moreover the additional
low-density condition allows us to consider the non barionic
hadronic matter as a pion gas.

At low temperatures most of the pion interactions occur at low
energies. The low energy pion interactions are completely
determined by the chiral symmetry pattern of the strong
interactions. In the chiral limit (vanishing $u$ and $d$ quark
masses) the strong interaction dynamics is invariant under the
$SU(2)_L\times SU(2)_R$ chiral transformations. However this
symmetry is supposed to be spontaneously broken to the diagonal or
isospin group $SU(2)_{L+R}$, being the  pions the corresponding
Goldstone bosons. In the real world quarks masses are different
from zero and then this symmetry breaking pattern is not exact
resulting in a  small value (when compared with other hadrons) for
the pion mass $M_\pi$. In any case chiral symmetry determines
completely the low energy pion interactions in terms of $M_ \pi$
and $F_\pi$ (Weinberg low energy theorems). The corrections
appearing at higher energies can be computed in a systematic way
by means of  Chiral Perturbation Theory \cite{ChPT}. In the chiral
limit the pion scattering amplitude goes to zero in the low energy
limit. Out of the chiral limit it is proportional to $M_\pi^2/
F_\pi^2$. In any case the interaction is small in this regime.
Therefore we arrive to the conclusion that non barionic hadronic
matter at low temperature and density can be described as a weakly
interacting pion gas. Another important observation is that at low
temperatures (energies) most of the pion collisions are elastic.
This fact implies in particular that the number of pions is
conserved. Thus it is possible to introduce the corresponding pion
chemical potential $\mu$ (not to be confused with the more usual
chemical potential associated to the barionic number). This makes
it possible to consider the pion gas at thermal equilibrium at
temperature $T$ for different values of the chemical potential
$\mu$, or what it is the same, different pion densities,  and not
only the case of chemical equilibrium corresponding to $\mu=0$.

According to the previous discussion the simplest state for non
barionic hadronic matter at low temperatures and densities is just
a gas of weakly interacting pions in thermal equilibrium which
could be described in a first approximation with a Bose-Einstein
distribution. In addition, at very low temperatures the average
pion energy is much smaller than the pion mass allowing for  to
have a non relativistic description of the pion gas. As discussed
later, the low density condition will prevent  the formation of
the Bose-Einstein condensate out of the pions, thus making the
description of the hadronic matter in the mentioned regime
especially simple.

More generally it is possible to consider the case when the
equilibrium is only local. In this case the temperature, density
(or pion chemical potential) and the overall velocity are
space-time functions. In this situation it is possible to have an
hydrodynamic description of the system as that considered in the
Bjorken picture of  heavy ion collisions. A departure from local
equilibrium will give rise to dissipative processes like
viscosity,  which is the main topic of this work. The evolution of
the distribution functions outside equilibrium can be studied by
using the Uehling-Uhlenbeck equation, which is the quantum version
of the Boltzmann transport equation. The computation of the
transport coefficients such as viscosity or heat conductivity
requires the solution of the transport equation for different
specific perturbations of the local equilibrium distribution.

As it was mentioned above, in this work we are interested in the
computation of the viscosity of the hadronic fluid with the
approximations discussed in the previous paragraphs. In fact it is
possible to have a good estimation of this magnitude by neglecting
quantum effects. In this case the equilibrium distribution
function is just the Boltzmann distribution instead the
Bose-Einstein distribution. Moreover  the elastic scattering pion
cross section is constant at low energies and as we will see it is
given by
\begin{equation}
\sigma=\frac{23}{384}\frac{M_\pi^2}{ \pi F_\pi^4}
\end{equation}
In addition, at low temperatures we can use the non relativistic
approximation so that our problem is equivalent to the classical
computation of the viscosity of a gas of hard spheres of radius
$R$ so that $\sigma= \pi R^2$. By using the well known result in
this case (see for instance \cite{Landau2}) we find  the viscosity
to be given by
\begin{equation}\label{eta}
\eta=\frac{1920}{368}\frac{F_\pi^4\pi^{3/2}}{M_\pi}\sqrt{
\frac{T}{M_\pi}}
\end{equation}
This formula provides a nice estimation of the viscosity of the
pion gas. However it does not take into account any quantum effect
that can be important at low temperatures. In addition it does not
include any dependence in the pion chemical potential, i.e. in the
pion density.

In the following we will show how these effects can be properly
taken into account to find the pion gas viscosity in terms of the
temperature and chemical potential (or density) in the low
temperature and density regime at the leading order. We will also
estimate the magnitude of the next to leading corrections. The
plan of the paper goes as follows: In sec.2 we review the main
equations of the non-relativistic ideal pion gas to set the
notation and for further reference. In sec.3 we describe briefly
the fundamentals of the kinetic theory needed to follow the
present work and in sec.4 we  show how the hydrodynamic equations
emerge from it. In sec.5 we obtain the equations for the transport
coefficients in terms of the appropriate solutions of the
linearized transport equations. In sec.6 we simplify these
equations to write them in useful form for finding explicit
formulae for the viscosity. In sec.7 we give the cross-section for
the low energy elastic pion scattering obtained from the lowest
order chiral perturbation theory. In sec.8 we obtain the lowest
order terms of the pion gas viscosity. In sec.9 we show our
numerical results and discuss the magnitude of the next to leading
corrections and in sec.10 we set the main conclusions of our work.
Finally in the appendix we study the properties of the polynomials
introduced in this work that play a role similar to the standard
Sonine polynomials in the more familiar classic computations.

\section{State equation}

In this section we review briefly the equation of state of a free
non-relativistic bosonic gas, which will provide our description
of the pion gas at low temperature and density. As it is well
known \cite{Amit}, for a large volume the pressure can be written
as
\begin{eqnarray}
P=&-&\frac{g T}{2\pi^2}\int^{\infty}_{0}dp\;p^2
\log\left[1-e^{\beta(\mu-E(p))}\right] \nonumber \\ &-&\frac{g
T}{V}\log\left[1-e^{\beta\mu}\right]
\end{eqnarray}
where $g$ is the number of pions $(g=3)$, $T=1/\beta$ is the
temperature, $V$ is the volume, $\mu$ is the pion chemical
potential and $E(p)$ is the energy of a non relativistic pions in
terms of the momentum i.e. $\vec p=M_\pi \vec v$, $E(p)=p^2/ 2
M_\pi$. The number density or the number of pions per unit of
volume is given by
\begin{equation}
n=\frac{g}{2\pi^2}\int_{0}^{\infty}dp\;p^2
\frac{1}{e^{-\beta(\mu-E(p))}-1}+\frac{g}{V}
\frac{1}{e^{-\beta\mu}-1}
\end{equation}
where the last term is the number of pions with zero momentum per unit
of volume. When this term is not negligible it is responsible for the
Bose-Einstein condensate. We thus define
\begin{equation}
n_0=\frac{g}{V}\frac{1}{e^{-\beta\mu}-1}
\end{equation}
Clearly the above equations make sense only when the chemical
potential satisfies $\mu\le 0$. As a consequence, in the
thermodynamic limit, where $N,V\rightarrow\infty$ with $N/V$
constant, we find two phases: In the first one (normal phase)
$\mu\le 0$ and $n_0=0$. However if we lower the temperature,
keeping a fixed density, $\mu$ increases until $\mu =0$. At that
point $n_0\neq 0$ and the ground state density starts to grow
forming the Bose-Einstein condensate.  Eventually, at $T=0$, all
pions are in the ground state so that $n=n_0$. The critical
temperature $T_c$ where the phase transition occurs can be found
to be
\begin{equation}
 T_c\equiv \frac{2\pi}{M_\pi}\left(
\frac{n}{ g \zeta_{3/2}}\right)^{2/3} \nonumber
\end{equation}
where $\zeta_{3/2} $ is the Riemann $\zeta$ function evaluated at
$3/2$. Therefore the critical temperature can be made arbitrary
small by lowering the density. In this work we are interested in
the the low temperature and the low density regime. In order to
simplify the analysis we will not consider the contribution of the
condensate. Thus for a given temperature $T$, we will always
assume that the density will be low enough so that $T>T_c$ and no
significant fraction of pions will have zero momentum (see
\cite{virial} for a discussion of the pion condensate).

In the following it will also be convenient to introduce the
fugacity
\begin{equation}
z\equiv\exp\left({\mu\over T}\right)\nonumber
\end{equation}
and then the above equations for the density and the pressure can
be written as
\begin{equation}
 n=\frac{g}{2\pi^2}\int_0^\infty  \frac{p^2
}{ z^{-1} e^\frac{p^2}{ 2M_\pi T}-1} d p
\end{equation}
and
\begin{equation}
 P=-\frac{g T }{
2\pi^2}\int_0^\infty p^2 \ln\left(1-ze^{-\frac{p^2}{2M_\pi T}}\right)d p
\end{equation}
Introducing the variable $x=p^2/(2M_\pi T)$ we have
\begin{equation}
  n=\frac{g
}{ 4\pi^2}\left(2M_\pi T\right)^{3/2}\int_0^\infty \frac{x^{1/2}}{ z^{-1}
e^x-1} d x
\end{equation}
and
\begin{equation}
 P=\frac{g}{6\pi^2}\left(2M_\pi T\right)^{3/2}T
\int_0^\infty \frac{x^{3/2} }{ z^{-1} e^x-1} d x
\end{equation}
where the  integrals can be written in terms of the $g_n$
functions \cite{Abramow} as
\begin{equation}
\int \frac{x^n }{ z^{-1} e^x-1} d x=\Gamma(n+1)
g_{n+1}(z)
\end{equation}
with
\begin{equation}
g_n(z)=\sum_{l=1}^\infty \frac{z^l}{ l^{n+1}} \nonumber
\end{equation}
so that
\begin{equation}
g_s(1)=\zeta_{s} \nonumber
\end{equation}
and
\begin{equation}
g_{s-1}(z)= z\frac{d g_s(z)}{ d z}\nonumber
\end{equation}
Therefore we have for a non relativistic free pion gas at
equilibrium  at low temperature and density
\begin{equation}
n=g\left(\frac{M_\pi T}{ 2\pi}\right)^{3/2}g_{3/2}\left(z\right)
\end{equation}
and
\begin{equation}
P=g\left(\frac{M_\pi T}{ 2\pi}\right)^{3/2}T g_{5/2}\left(z\right)
\end{equation}
Thus the equation of state can be written as
\begin{equation}
P=nT\frac{g_{5/2}\left(z\right)}{g_{3/2}\left(z\right)}
\end{equation}

\section{The kinetic theory for pions}

The possibility of having an hydrodynamic description of the pion
gas requires the definition of macroscopic fields such as the
velocity $\vec V=\vec V\left(\vec r,t\right)$, the pressure
$P=P\left(\vec r,t\right)$, the number density $n=n\left(\vec
r,t\right)$, the temperature $T=T\left(\vec r,t\right)$ and
others. In other words, it is needed to have a large number of
pions in each volume element $dV$. The statistical description of
the gas is based on the (one-particle) distribution function
$f\left(\vec r, \vec v, t\right)$ which gives the number of pions
inside the phase-space volume element $d\vec r\  d\vec v$ at the
instant $t$. In fact the phase-space volume element is given by
$d\vec r\  d\vec p$ but in the case considered here the difference
amounts to a constant factor which we include in the
definition of the distribution function $f$. Thus the total number
of pions will be given by
\begin{equation}
N=\int f \left(\vec r,\vec v, t\right)d \vec r d \vec v=\int
n\left(\vec r,t\right)d \vec r
\end{equation}
As we are interested here in the behaviour of the pion gas at low
temperatures and densities, most of the the pion interactions will
occur at low energies. Then the pion-pion interactions will be
essentially elastic and the total number of pions will be a
conserved quantity. This in particular means that it makes sense
to introduce the chemical potential $\mu=\mu\left(\vec r,t\right)$
related with the pion number even in presence of low-energy pion
interactions. The evolution of the distribution function is
determined by the BBKGY hierarchy (from N.N. Bogoliubov, M. Born,
G. Kirkwood, H.S. Green and J. Yvon) \cite{Liboff}. This is a set
of $N$ coupled equations for the $n$-particle distribution
functions $f_{(n)}\left(\vec r_1,\vec v_1,\vec r_2,\vec
v_2,...,\vec r_n,\vec v_n,t\right)$ with $f=f_{(1)}$. However, for
low-density gases it can be a good approximation to truncate the
BBKGY hierarchy. For example by using the Bogoliubov ansatz, i.e.
by writing the two-particle function in terms of an appropriate
product of two one-particle functions (molecular chaos
hypothesis), it is possible to decouple the first BBKGY equation
from the others to obtain the well known Boltzmann equation. This
equation was modified by Uehling and Uhlenbeck in order to include
the quantum statistic effects (see for example \cite{UehlingU}).
For the bosonic case the Uehling-Uhlenbeck equation reads
\begin{equation}
\label{estado1}
\left(\partial_t+\vec v_1 \vec \nabla_{\vec r}\right)f_1=C\left[f_1\right]
\end{equation}
where
\begin{eqnarray}
\label{colision1}
C\left[f_1\right]=\int d \vec v_2 d \sigma u
\left[f'_1f'_2\left(1+\frac{f_1}{A}\right)\left(1+\frac{f_2}{
A}\right)\right.\nonumber \\ \left. - f_1f_2\left(1+\frac{f'_1}{
A}\right)\left(1+\frac{f'_2}{ A}\right)\right]
\end{eqnarray}
is the  collision term or collision functional. The above equation
describes the irreversible space-time evolution of the (one-point)
distribution function $f$, being this irreversible behaviour a
consequence of the truncation of the BBKGY hierarchy. As usual we
use for short the notation $f_i=f\left(\vec r_i,\vec v_i,t\right)$
and $f_i'=f\left(\vec r_i',\vec v_i',t\right)$ with $i=1,2$.
$d\sigma$ is the differential cross-section corresponding to the
elastic scattering of two pions with initial and final velocities
$\vec v_1, \vec v_2$ and $\vec v _1', \vec v_2'$, respectively.
$u$ is the relative velocity of the initial particles $u=\mid \vec
v_1- \vec v_2 \mid$ and, finally, the normalization constant $A$
is given by  $A=gM_\pi^3/\left(2\pi\right)^3$.

As it is the case of the Boltzmann equation, the Uehling-Uhlenbeck
equation typically drives the gas from arbitrary initial distributions
to the equilibrium distribution $f^0$ which is a fixed point of this
equation in the sense that the collision functional vanishes on it
\begin{equation}
C\left[f^0\right]=0
\end{equation}
For a gas moving at a macroscopic velocity $\vec V$ the equilibrium
distribution corresponding to temperature $T$ and pion chemical
potential $\mu$ is given by
\begin{equation}
f^0\left(\vec r, \vec v,t\right)=A
\left(e^{\frac{1}{T}\left(M_\pi\frac{\left(\vec v-\vec
V\right)^2}{2}-\mu\right)}-1\right)^{-1}
\end{equation}
from which the properties of free pion gas given in the previous
section can be derived. The case of local equilibrium is also
described by this distribution but having the $\vec V, T$ and $\mu$
parameters depending on the position and time. This case corresponds
to an ideal fluid. However, in order to describe dissipative
processes, such as viscosity or thermal conduction, it is needed to
consider the case where the pion gas is not even in local thermal
equilibrium.

\section{Hydrodynamic equations}

Given some magnitude $\psi$ it is possible to compute its statistic
average by using the distribution function as
\begin{equation}
\left<\psi\right>=\frac{\int f \psi d\vec r}{\int f d \vec r}=\frac{\int f
\psi d \vec r}{n\left(\vec r,t\right)}
\end{equation}
where $n=n\left(\vec r,t\right)$ is the pion number density. This
average can be understood as the corresponding macroscopic magnitude
in the hydrodynamic approach \cite{Balescu}. For example, the
macroscopic fluid velocity $\vec V$ is given by
\begin{equation}
\vec V=\frac{\int f \vec v d\vec v}{\int f d \vec v}=\frac{1}{n}
\int f v d \vec v
\end{equation}
The equation governing the evolution of any of these magnitudes
(transport equations) can be obtained by multiplying $\psi_1$ by the
Uehling-Uhlenbeck equation and then integrating $\vec v_1$ so that
\begin{widetext}
\begin{equation}
\label{21}
\int d\vec v_1
\left(\partial_t+\vec v_1 \vec \nabla_{\vec r}\right)f_1\psi_1=
\int d \vec v_1 d \vec v_2 d \sigma u \psi_1  \left(f'_1f'_2
\left(1+A^{-1}f_1\right)\left(1+A^{-1}f_2\right) -
f_1f_2\left(1+A^{-1}f'_1\right)\left(1+A^{-1}f'_2\right)\right)
\end{equation}
The r.h.s. term can be written in a more symmetric fashion as
\begin{equation}
\frac{1}{4} \int d\vec v_1 d\vec v_2 d\vec v_3 d\vec v_4 \mid T \mid
^2 \left(\psi_1+\psi_2-\psi_1'-\psi_2'\right)\left(f'_1f'_2
\left(1+A^{-1}f_1\right)\left(1+A^{-1}f_2\right) -
f_1f_2\left(1+A^{-1}f'_1\right)\left(1+A^{-1}f'_2\right)\right)
\end{equation}
where $ \mid T \mid ^2 $ is defined through the relation
\end{widetext}
\begin{equation}
d\sigma=\frac{\mid T \mid ^2}{\mid \vec v_1 -\vec v_2    \mid}d
\vec v_1' d\vec v_2'
\end{equation}
From this form of the collision terms it becomes apparent than it
vanishes for any quantity $\psi$ which is conserved in the collision
such as the total energy or momentum. On the other hand, by using the
definition of statistical average, the first term of (\ref{21}) can be
written as:
\begin{equation}
\partial_t n \left<\psi\right>+\vec \nabla_{\vec r}n\left<\psi \vec
v\right>
\end{equation}
Thus for conserved magnitudes this term vanishes. In order to find
the basic hydrodynamic equations it is useful to write the
velocity as $\vec v=\vec V + \vec \xi$, i.e. $\vec \xi$ is the
pion velocity relative to the macrocopic velocity $\vec V$ and
therefore it averages to zero. Choosing $\psi=1, \vec v$ and $v^2$
(corresponding to the conservation of a constant, the mumentum and
the energy in the elastic collision) we obtain the equations
\begin{widetext}
\begin{eqnarray}
\psi &=&1\longrightarrow \partial_t n+ \vec \nabla n \vec V=0
\nonumber \\
\psi &=&\vec v \longrightarrow \partial_t n V_i
+\nabla_j n V_i V_j + \nabla_j n \left<\xi_i \xi_j\right> =0  \\
\psi &=& v^2 \longrightarrow
\partial_t n \left(V^2 + \left<\xi^2\right>\right) + \nabla_i n V_i
\left(V^2 +\left<\xi^2\right>\right)
+ \nabla_i 2 n V_j \left<\xi_i \xi_j\right> + \nabla_i n\left<\xi_i
\xi^2\right>=0  \nonumber
\end{eqnarray}
\end{widetext}
In order to understand the meaning of the above equations we can
introduce the following macroscopic quantities:
\begin{equation}
P_{ij}=M_\pi n \left<\xi_i\xi_j\right>
\end{equation}
is the pressure tensor,
\begin{equation}
Q=\frac{1}{2}M_\pi n \left<\xi^2\right>
\end{equation}
is the internal energy density and
\begin{equation}
q_{i}=\frac{1}{2}M_\pi n \left<\xi_i\xi^2\right>
\end{equation}
is the energy (heat) flux. Due to the isotropy of the pion fluid
the pressure tensor can be taken as diagonal, i.e.
$P_{ij}=P\delta_{ij} $. In this case $Q=3P/2$ and $q_i=0$. Thus,
(by taking $\vec V=0$ but not its derivatives, which can always be
done) we arrive to the standard ideal fluid equations, namely the
continuity equation
\begin{equation}
 \partial_t n+n\nabla_iV_i=0,
\end{equation}
the Euler equation
\begin{equation}
n M_\pi \partial_t V_i+\nabla_i P=0
\end{equation}
and the energy conservation equation
\begin{equation}
\partial_t Q+ \frac{5}{3}Q\nabla_i V_i=0
\end{equation}

In spite of the above equations for the ideal fluid, it is well
established that an irreversible flux of energy and momentum
appear in an inhomogeneous gas. This flux gives rise to the well
known dissipative processes of thermal and conduction and
viscosity. For example an small gradient of temperature produces a
heat flow which in a first approximation can be written as
\begin{equation}
q_i=-\kappa \nabla_i T
\end{equation}
where $\kappa$ is a coefficient called thermal conductivity. In
order to include momentum diffusion effects we add an extra term
to the isotropic pressure tensor:
\begin{equation}
P_{ij}=P\delta_{ij}+P'_{ij}
\end{equation}
The requirement for this tensor to vanish for uniform translations
and rotations of the fluid, makes it possible to write it to first
order in the velocity derivatives as
\begin{equation}
P'_{ij}=-\eta \left(\frac{\partial V_i}{\partial x_j}+\frac{\partial
V_j}{\partial x_i}-\frac{2}{ 3}\sum_{l=1}^3 \frac{\partial V_l}{
\partial x_l}\delta_{ij}\right)-\zeta \sum_{l=1}^3 \frac{\partial V_l }{ \partial
x_l}\delta_{ij}
\end{equation}
Thus it is needed to introduce two coefficients $\eta$ and $\zeta$
which are usually called first and second viscosity coefficients.
The introduction of the dissipative effects does not modify the
continuity equation. However the Euler equation becomes
\begin{widetext}
\begin{equation}
\partial_t n V_i +\nabla_j n V_i V_j + \frac{1}{ M_\pi}\nabla_j
P\delta_{ij}= \frac{1}{ M_\pi}\nabla_j\left[\eta
\left(\frac{\partial V_i}{ \partial x_j}+\frac{\partial V_j}{ \partial
x_i}-\frac{2}{ 3}\sum_{l=1}^3 \frac{\partial V_l }{ \partial
x_l}\delta_{ij}\right)+\zeta \sum_{l=1}^3 \frac{\partial V_l }{
\partial x_l}\delta_{ij}\right]
\end{equation}
which is known as the Navier-Stokes equation. The energy
conservation equation is in this case
\begin{eqnarray}
 \partial_t \left(nV^2 +\frac{2}{ M_\pi} Q\right) + \nabla_i V_i
  \left(nV^2 +\frac{2}{
M_\pi} Q\right) + \nabla_i \frac{2}{ M_\pi} V_jP\delta_{ij} = \nonumber\\
\nabla_i \kappa \nabla_i T+\frac{2}{ M_\pi}\nabla_i V_j\left[\eta
\left(\frac{\partial V_i}{ \partial x_j}+\frac{\partial V_j}{ \partial
x_i}-\frac{2}{ 3}\sum_{l=1}^3 \frac{\partial V_l}{ \partial
x_l}\delta_{ij}\right)+\zeta \sum_{l=1}^3 \frac{\partial V_l}{
\partial x_l}\delta_{ij}\right]
\end{eqnarray}
\end{widetext}
Thus the description of the dissipative flow of energy and
momentum requires the introduction of three transport coefficients
$\kappa, \eta$ and $\zeta$. However, for an incompressible fluid
the second (shear) viscosity does not play any role. In the
following we will assume that this is the case for the pion gas
considered here.

\section{Computation of the transport coefficients}

As it has been mentioned above the description of the dissipative
processes requires going beyond local equilibrium. Thus, in order
to compute the transport coefficients we have to consider a
distribution function $f$ slightly perturbed from the equilibrium
distribution $f^0$, i.e.
\begin{equation}
\label{ligin}
f=f^0+\delta f=f^0+f^0\frac{\chi}{T}
\end{equation}
where $\chi$ is an arbitrary function of the velocities which
represents the inhomogeneous contribution to the distribution
function. This contribution will be assumed to be small in the
sense that $\chi /T \ll 1$, since we are interested only in the
computation of the transport coefficients. The $\chi$ function
must by determined by solving the Uehling-Uhlenbeck equation which
can be linearized with respect to $\chi$. Then it is possible to
compute the transport coefficients from this function. For
example,  the heat flux can be written as
\begin{equation}
q_i=\frac{1}{ 2}M_\pi\int f^0\left(1+{\chi\over T}\right)v_iv^2 d \vec v
\end{equation}
Taking into account that this flux vanishes for an homogeneous gas
and using the definition of the heat conductivity we have
\begin{equation}
-\kappa\nabla_i T=\frac{M_\pi}{ 2T}\int f^0\chi v_iv^2 d \vec v
\end{equation}
Now it is useful to write the $\chi$ function as
\begin{equation}
\chi=-\nabla_i T h_i\left(\vec v\right)=-\nabla_i T v_i h\left(v\right)
\end{equation}
Therefore
\begin{equation}
\kappa \nabla_i T= \kappa_{ij}\nabla_j T
\end{equation}
with the tensor conductivity $\kappa_{ij}$ being given by
\begin{equation}
\kappa_{ij}\equiv\frac{M_\pi}{ 2T}\int f^0 h v_jv_iv^2 d \vec v
\end{equation}
For an isotropic gas $\kappa_{ij}=\kappa \delta_{ij}$ so that, we
have the following equation for the heat conductivity
\begin{equation}
\kappa=\frac{M_\pi}{ 6T}\int f^0 h v^4 d \vec v
\end{equation}
In a similar way it is possible to write the viscosity in terms of
an $h$ function. In order to do that we write the pressure tensor
in terms of the distribution function
\begin{equation}
P_{ij}=M_\pi\int f^0\left(1+\frac{\chi}{ T}\right)v_iv_j d \vec v
\end{equation}
For the present calculation it is enough to consider the case
$\vec V=0$ (but not its derivatives). Then the non isotropic
component of the pressure tensor can be written as
\begin{equation}
P'_{ij}=\frac{M_\pi}{T}\int f^0 \chi v_iv_j d \vec v
\end{equation}
On the other hand for an incompressible fluid this tensor is
\begin{equation}
P'_{ij}=-2\eta\left(V_{ij}-\frac{1}{ 3}\nabla_hV_h\delta_{ij}\right)
\end{equation}
where we have introduced $V_{ij}= (\partial_i V_j$ $+\partial_j
V_i)/2$. For further convenience we choose the perturbation to
have the form
\begin{equation}
\chi=-\left(V_{kl}-\frac{1}{ 3}\nabla_h V_h
\delta_{kl}\right)h_{kl}\left(\vec v\right)
\end{equation}
where $h_{kl}$ is a microscopic velocity dependent quantity which can
be written as
\begin{equation}
 h_{kl}=\left(v_k v_l-\frac{1}{ 3}v^2 \delta_{kl}\right)h\left(v\right)
\end{equation}
Therefore, by equating the two forms of $P'_{ij}$ we find
\begin{equation}
2 \eta \left(V_{ij}-\frac{1}{ 3}\nabla_h V_h \delta_{ij}\right)=
\eta_{ijkl}\left(V_{kl}-\frac{1}{ 3}\nabla_h V_h \delta_{kl}\right)
\end{equation}
where $\eta_{ijkl}$ (the viscosity tensor) is given by
\begin{equation}
\eta_{ijkl}\equiv{M_\pi }{T}\int v_iv_j\left(v_kv_l-\frac{1}{3}v^2
\delta_{kl}\right)f^0\left(v\right) h\left(v\right) d \vec v
\end{equation}
Due to the several symmetries of this tensor it can written as
\begin{equation}
\eta_{ijkl}=\eta \left(\delta_{ik}\delta_{jl}+\delta_{il}\delta_{jk}-
\frac{2}{ 3}\delta_{ij}\delta_{kl}\right)
\end{equation}
so that we have
\begin{equation}
\eta=\frac{M_\pi }{ 10T}\int v_i v_j\left(v_iv_j-\frac{1}{
3}v^2\delta_{ij}\right) f^0\left(v\right)h\left(v\right)d \vec v
\end{equation}
As the $h$ function depends only on the velocity modulus it is
possible to perform the angular integrations
\begin{equation}
\eta= \frac{4\pi }{ 15T}A \int_0^\infty \frac{1}{
z^{-1}\exp\left(\frac{M_\pi v^2 }{ 2T}-1\right)} v^6 h\left(v\right) d \vec v
\end{equation}
In the following it will be useful to consider $h$ as a function of
the adimensional variable $x$, i.e. $h=h\left(x\right)$ defined as
\begin{equation}
x=\frac{M_\pi v^2}{2T}
\end{equation}
At this point it is customary to develop the $h\left(x\right)$ in
terms of the Sonine polynomials. However it is more appropriate in
our case to introduce a new family of orthogonal polynomials
$P_r^s\left(z;x\right)$ since we are considering the
Uehling-Uhlenbeck as the transport equation (i.e. we have taken
into account quantum effects) instead of the more common Boltzmann
equation. The definition and the main properties of this new
polynomials can be found in appendix A. In terms of $P_r^s$ the
$h$ function can be written as
\begin{equation}
 h\left(x\right)=\sum_{s=0}^\infty B_s P_r^s\left(z;x\right)
\end{equation}
Therefore
\begin{equation}
\eta=\frac{M_\pi }{10T}\left(\frac{2T}{ M_\pi}\right)^{5/2}A\int_0^\infty d
x\frac{x^{5/2}}{ z^{-1}e^x-1} \sum_{s=0}^\infty B_s P_r^s\left(z;x\right)
\end{equation}
and the distribution function as
\begin{eqnarray}
f= f^0+\frac{f^0}{ T}\left(V_{ij}-\frac{1}{ 3}\delta_{ij}V_{hh}\right)
\left(v_iv_j-\frac{1}{ 3}\delta_{ij}v^2\right)\nonumber\\
\sum_{s=0}^\infty B_s
P_r^s\left(z;\frac{M_\pi}{ 2T}v^2\right)
\end{eqnarray}
For all the families of polynomials, i.e., for all the $r$ values,
we have exact conservation of the collision invariants. This fact
makes possible to choose $r$ so that we get the simplest integral.
The appropriate value turns out to be $r=5/2$ and then we get
\begin{equation}
\eta=\frac{A\pi^{3/2}}{ 2}\left(\frac{2T}{ M_\pi}\right)^{5/2}g_{
7/2}\left(z\right) B_0
\end{equation}
In this way the computation of the viscosity of the pion gas has
been reduced to the computation of the $B_0$ coefficient.

\section{Simplifying the Uehling-Uhlenbeck equation}

As discussed above the the perturbed distribution function must be
a solution of the Uehling-Uhlenbeck equation (\ref{estado1}). In
order to solve this equation it is quite convenient to write the
l.h.s. in terms of macroscopic quantities, as the temperature,
velocity, density and fugacity, and their derivatives. This can be
done by using the equation of state and the hydrodynamic equations
as follows: First of all at lowest order we have
\begin{widetext}
\begin{equation}
\partial_t f+\vec v \vec \nabla f\simeq
\partial_t f^0+\vec v \vec \nabla f^0=
A\partial_t \left(\frac{1}{ z^{-1} \exp\left(\frac{E}{
T}\right)-1}\right) +A\vec v \vec
\nabla \left(\frac{1}{ z^{-1} \exp\left(\frac{E}{ T}\right)-1}\right)
\end{equation}
Thus the temporal derivative is
\begin{equation}
\partial_t f^0=-\frac{{f^0}^2z^{-1}\exp\left(\frac{E}{ T}\right)}{ A}
\left(\frac{1}
{ T}\partial_t E -
\frac{E}{ T^2}\partial_t T -\partial_t \log z\right)
\end{equation}
where the energy $E$ is given by
\begin{equation}
E=\frac{M_\pi}{ 2}\left(\vec v -\vec V\right)^2
\end{equation}
so that we have
\begin{equation}
\partial_t f^0=\frac{{f^0}^2}{A z\exp\left(-\frac{E}{ T}\right)}
\left(\partial_t \log
z+\frac{E}{ T^2}\partial_t T+\frac{M_\pi}{ T}v_i \partial_t V_i\right)
\end{equation}
In a similar way we can obtain
\begin{equation}
v_j\nabla_j f^0=\frac{{f^0}^2v_j}{ A z\exp\left(-\frac{E}{ T}\right)}
\left(\nabla_j \log
z+\frac{E}{ T^2}\nabla_j T+ \frac{M_\pi}{ T}v_i \nabla_j V_i\right)
\end{equation}
Thus we have
\begin{equation}
\partial_t f^0 +v_j \nabla_j f^0=
\frac{{f^0}^2}{ A z\exp\left(-\frac{E}{ T}\right)}\left(\partial_t \log z+\frac{E}{
T^2}\partial_t T+\frac{M_\pi}{ T}v_i \partial_t V_i+v_j\nabla_j
\log z+\frac{E}{ T^2}v_j\nabla_j T+\frac{M_\pi}{ T} v_i v_j
V_{ij}\right)
\end{equation}
Here it is possible to use the state equation for the free pion
gas and the ideal fluid equations (Euler, continuity and energy
conservation) to find
\begin{equation}
\partial_t f^0 +v_j \nabla_j f^0=\frac{{f^0}^2}{ A z\exp\left(-\frac{E}{
T}\right)}\left[\frac{1}{ T}\left(\frac{E}{ T}-\frac{5}{ 2}
\frac{g_{5/2}\left(z\right)}{ g_{3/2}\left(z\right)}\right)v_i\nabla_i
T+\left(\frac{M_{\pi}}{ T}v_iv_j-\frac{2}{ 3}\frac{E}{ T}
\delta_{ij}\right)V_{ij}\right]
\end{equation}
On the other hand, substituting expression (\ref{ligin}) into
(\ref{colision1}), expanding to first order in the perturbation
$\chi$ and taking benefit from the equilibrium distribution
function relation
\begin{equation}
{f_1^0}'  {f_2^0}' \left(1+A^{-1} f_1^0\right) \left(1+ A^{-1}f_2^0\right)=
f_1^0 f_2^0 \left(1+ {f_1^0}'\right)\left(1+A^{-1} {f_2^0}'\right)
\end{equation}
we obtain
\begin{eqnarray}
\label{primero}
f_1'  f_2'\left( 1+A^{-1}  f_1\right)\left(1+A^{-1} f_2\right)-
f_1  f_2\left( 1+A^{-1}  f_1'\right)\left( 1+A^{-1}  f_2'\right) =
\nonumber \\
{A^2\over T}{z^{-2}e^{E/T}\over \left(z^{-1}e^{E_1/T}-1\right)\left(z^{-1}
e^{E_2/T}-1\right)\left(z^{-1}e^{E'_1/T}-1\right)
\left(z^{-1}e^{E'_2/T}-1\right)}\Delta\left[\chi\left(1-z\exp\left(-\frac{E}{
T}\right)\right)\right]
\end{eqnarray}
where $E=E_1+E_2=E_1'+E_2'$ and the $\Delta$ symbol is defined as
\begin{equation}
\Delta\left[f\left(x\right)\right]\equiv f
\left(x_1'\right)+f\left(x_2'\right)-f\left(x_1\right)-f\left(x_2\right) \
\end{equation}
Thus the whole collision term can be written as
\begin{equation}
C\left[f_1\right]=
\frac {1}{ A^2T}\int d \vec v_2 d \sigma\ u\
z^{-2}\exp\left({E/T}\right)f_1f_2f_1'f_2'\Delta
\left[\chi\left(1-z\exp\left(-\frac{E}{
T}\right)\right)\right]
\end{equation}
where we have omitted the superindex $0$ in the equilibrium
distribution function. In this way the complete transport equation
for the slightly unhomogeneous pion gas becomes
\begin{eqnarray}
\frac {f_1^2} { ze^{-\frac{M_{\pi}v^2_1}{ 2T}  }  }
\left[\left(\frac{M_\pi v^2_1}{ 2T}-\frac{5}{ 2}\frac
{g_{5/2}\left(z\right)}{
g_{3/2}\left(z\right)}\right)v_1{}_i\nabla_i
T+M_\pi\left(v_1{}_iv_1{}_j -\frac{1}{
3}v^2_1 \delta_{ij}\right)V_{ij}\right]=     \nonumber
\\ \frac{1}{ A}\int
d \vec v_2 d \sigma\ u\ z^{-2}e^{\frac{M_\pi}{ 2T}\left(4U^2+u^2\right)}
f_1f_2f_1'f_2'\Delta\left[\chi\left(1-ze^{-\frac{M_\pi v^2}{ 2T}}\right)\right]
\end{eqnarray}
\end{widetext}
where we have defined $\vec U=\left(\vec v_1+\vec
v_2\right)/2=\left(\vec v_1'+\vec v_2'\right)/2$, i.e. it is the
collision center of mass velocity. Once again $\vec u= \vec
v_1-\vec v_2$ is the relative velocity of the incident pions.
Thus, in particular, we have
\begin{equation}
E=\frac{M_\pi}{ 2}\left(2U^2+\frac{1}{ 2}u^2\right)
\end{equation}

\section{Low-energy pion cross-section}

In order to solve the above transport equation it is needed to
know the differential cross section for pion scattering. As we are
interested in the low energy regime the most appropriate approach
is the Chiral Perturbation Theory which relies in the chiral
symmetry pattern of the strong interactions. In particular it is
quite useful the chiral Lagrangian approach suggested by Weinberg
and extended up to the one loop level by Gasser and Leutwyler.
This chiral Lagrangian approach consists in a systematic expansion
on the pion field derivatives and the pion mass which are
considered to be of the same order. From the lowest order of this
expansion, the pion elastic scattering amplitude is given by
\begin{widetext}
\begin{equation}
T_{a_1,a_2,a_1',a_2'}\left(s,t,u\right)=\frac{s-M_\pi^2}{
F_\pi^2}\delta_{a_1 a_2} \delta_{a_1' a_2'} +\frac{t-M_\pi^2}{
F_\pi^2}\delta_{a_1 a'_1}\delta_{a_2 a_2'} +\frac{u-M_\pi^2}{
F_\pi^2}\delta_{a_1 a_2'}\delta_{a_2 a_1'}
\end{equation}
\end{widetext}
where the standard Mandelstan variables $s$, $t$ and $u$ are related
by $s+t+u=4 M_\pi^2$, the $a$ subindices refer to the pion isospin and
$F_{\pi}$ is the pion decay constant. In the total isospin basis there
are three independent amplitudes corresponding to $I=0,1,2$ which are
given by
\begin{eqnarray}
T^0 & = & \frac{2s-M_\pi^2}{ F_\pi^2} \nonumber\\
 T^1 & =  & \frac{t-u}{
F_\pi^2}                 \\              \nonumber
 T^2  & =  &
\frac{2M_\pi^2-s}{ F_\pi^2}
\end{eqnarray}
The averaged modulo squared amplitude is given by
\begin{equation}
\left|T\right|^2=\frac{1}{ \sum_I \left(2I+1\right)}\sum_I
\left(2I+1\right)\left|T^I\right|^2
\end{equation}
which leads to
\begin{equation}
\left|T\right|^2=\frac{1}{
9F_\pi^4}\left(21M_\pi^4+9s^2-24M_\pi^2s+3\left(t-u\right)^2\right)
\end{equation}
The relativistic differential cross-section for elastic pion
scattering is
\begin{equation}
d \sigma =\frac{\left|T\right|^2}{4uE_1 E_2} d LIPS
\end{equation}
where the differential Lorentz Invariant Phase Space is given by
\begin{eqnarray}
d LIPS  \equiv  C \left(2\pi\right)^4 \delta^4 \left(p_1+ p_2 - {p'_1}-
{p'_2}\right)\nonumber \\
\frac{d \vec {p'_1} }{ 2\left(2\pi\right)^3 E'_1} \frac{d \vec {p'_2}
}{ 2\left(2\pi\right)^3 E'_2}
\end{eqnarray}
where the constant $C=1/2$ takes into account the identity of the
final pions.

As discussed above we are interested in the non relativistic
regime where $\vec p_i=M_\pi\vec v_i$, $\vec P=2 M_\pi \vec U$ and
$E_i=M_\pi+ M_\pi v^2_i/2$. In this limit it is convenient to
write the cross section in terms of the center of mass and
relative velocities $\vec U$, $\vec u$ and $\vec u'$ defined as
\begin{equation}
\vec u'=\vec v'_1-\vec v'_2 \qquad u=u'
\end{equation}
After some standard calculations the non relativistic reduction of
the cross section can be found to be
\begin{widetext}
\begin{equation}
\label{sigma} d \sigma=\frac{23}{ 768}\frac{M_\pi^2}{\pi
F_\pi^4}\left(1-\frac{3}{ 2}U^2+\frac{13}{
184}u^2+U^2\cos\theta'\frac{2U+u \cos\theta'}{ 2U\cos\theta'+u}\right)d
\cos \theta'
\end{equation}
where $\theta'$ is the angle between by $\vec U$ and $\vec u'$.
\end{widetext}

\section{Lowest order viscosity for the pion gas }

As we have seen in previous sections the perturbation of the
distribution function appropriate for the computation of the pion gas
viscosity can be written as
\begin{eqnarray}
\chi=-\left(V_{ij}-\frac{1}{ 3}\nabla_h V_h\delta_{ij}\right) \left(v_i
v_j-\frac{1}{ 3}v^2\delta_{ij}\right)\nonumber \\ \sum_{s=0}^{\infty}B_s
P_{5/2}^s\left(z;x\right)
\end{eqnarray}
where the $B_s$ coefficients must be obtained by solving the
linerized transport equation being $B_0$ the dominant one. In
order to do so let us consider for a while a tensor $l_{kl}$ which
is an arbitrary function on $z$ and $\hat v_1$. For further
convenience we have defined the hat on any velocity  as
\begin{equation}
\hat v = \sqrt{\frac{M_\pi}{2 T}}v
\end{equation}
By multiplying this tensor by the transport equation and integrating
on $v_1$ we have
\begin{widetext}
\begin{eqnarray}
\label{ultima}
\int d \vec v_1 l_{kl} \left(z;\hat v_1\right) \frac{f_1^2}{
ze^{-{\hat v_1}^2}} \left[\left({\hat v_1}^2- \frac{5}{ 2}
\frac{g_{5/2}\left(z\right)} { g_{3/2}\left(z\right)}\right)
v_1{}_i\nabla_i T +M_\pi \left(v_1{}_i v_1{}_j-\frac{1}{ 3}v^2_1
\delta_{ij}\right) V_{ij}\right] = \nonumber \\
-V_{ij}\sum_{s=0}^\infty B_s\frac{1}{ A} \int d \vec v_1 d \vec v_2 d
\sigma\ l_{kl}\left(z;\hat v_1\right) u\ z^{-2} e^{\left(2 {\hat U}^2+
\frac{1}{ 2}{\hat u}^2\right)} f_1 f_2 f'_1 f'_2 \Delta\left[\left(v_i
v_j-\frac{1}{ 3}v^2\delta_{ij}\right)\left(1-ze^{-{\hat
v_1}^2}\right)P_{5/2}^s\left(z;{\hat v}^2\right)\right]
\end{eqnarray}
\end{widetext}
where we have assumed the divergence of the velocity to vanish since
the pion gas is taken to be incompressible. Now we choose the tensor
to be
\begin{equation}
l_{kl}\left(z;\hat v\right)= \left(v_kv_l-\frac{1}{
3}v^2\delta_{kl}\right) \left(1-ze^{-{\hat
v}^2}\right)P_{5/2}^t\left(z;{\hat v}^2\right)
\end{equation}
Then, one of the two integrals appearing in the r.h.s. of the
equation (\ref{ultima}) vanishes, namely
\begin{equation}
\int d \vec v_1 l_{kl} \left(z;\hat v_1\right) \frac{f_1^2}{
ze^{-{\hat v_1}^2}} \left({\hat v_1}^2- \frac{5}{ 2}
\frac{g_{5/2}\left(z\right)} { g_{3/2}\left(z\right)}\right)
v_1{}_i\nabla_i T=0
\end{equation}
Therefore, the two sides of the equation (\ref{ultima}) are
proportional to $V_{ij}$ which is arbitrary. Thus one possible
solution to the equation is found by eliminating this tensor in
both sides and then by contracting with the tensors
$\delta_{ik}\delta_{jl}$ so that the r.h.s. of the equation
(\ref{ultima}) becomes
\begin{eqnarray}
A^2 V_{ij}\int d \vec v \ \frac{M_\pi\left(v_kv_l-\frac{1}{ 3}v^2
\delta_{kl}\right)^2}{z^{-1}e^{{\hat v}^2}-1} \
P_{5/2}^t\left(z;{\hat v}^2\right)=\nonumber \\ V_{ij} A^2
\frac{5\pi^{3/2}M_\pi}{ 2}\left(\frac{2T}{
M_\pi}\right)^{7/2}g_{7/2}\left(z\right)\delta_{0t}
\end{eqnarray}
In order to compute the $B_s$ coefficients we have to proceed in a
similar way with the r.h.s. of the equation (\ref{ultima}). For
that purpose we will take the low energy limit of the differential
cross section which is just
\begin{equation}
\label{sigma0}
 d\sigma=\frac{23 M_\pi^2}{ 768\pi F_\pi^4}d \cos \theta'
\end{equation}
where $\theta'$ belongs to the interval $(0,\pi)$. Therefore, at
very low energies the elastic pion scattering cross section can be
considered as a constant, so that it is formally equivalent to the
elastic scattering of rigid spheres of radius
\begin{equation}
R=\frac{M_\pi}{ \pi F_\pi^2}\sqrt{\frac{23}{ 384}}
\end{equation}
After the manipulations described below the r.h.s of the transport
equation can be written the collision term as
\begin{equation}
V_{ij}\frac{23 M^2_\pi A^3 }{ 768F_\pi^4}\left(\frac{2T}{
M_\pi}\right)^{11/2}\sum_{s=0}^\infty B_s b_{st}\left(z\right)
\end{equation}
where we have defined
\begin{widetext}
\begin{eqnarray}
\label{bdefinition} b_{st}\left(z\right)&=&\int_0^\infty d U d u\int_0^\pi d
\theta' d \theta \int_0^{2\pi} d \phi \sin \theta' \sin \theta\ U^2
u^3 z^{-2} e^{2U^2+\frac{1}{ 2}u^2} \prod_{a_1,a_2,a'_1,a'_2}
\left(z^{-1}e^{v^2_a}-1\right)^{-1} \nonumber \\&& \Delta\left[\left(v_k
v_l-\frac{1}{ 3}\delta_{kl}
v^2\right)\left(1-ze^{-v^2}\right)P_{5/2}^s\left(z;v^2\right)\right]
\Delta\left[\left(v_kv_l-\frac{1}{ 3}\delta_{kl}
v^2\right)\left(1-ze^{-v^2}\right)P_{5/2}^t\left(z;v^2\right)\right]
\end{eqnarray}
\end{widetext}
Then the transport equation (\ref{ultima}) can be written as
\begin{equation}
\frac{23M_\pi A }{ 1920 \pi^{3/2}F_\pi^4}\left(\frac{2T}{ M_\pi}\right)^2
g^{-1}_{7/2}\left(z\right)\sum_{s=0}^\infty B_s b_{st}=\delta_{0t}
\end{equation}
Therefore by defining
\begin{equation}
{\cal B}_s\equiv\frac{23M_\pi A }{
1920\pi^{3/2}F_\pi^4}\left(\frac{2T}{ M_\pi}\right)^2
g^{-1}_{7/2}\left(z\right)B_s
\end{equation}
the viscosity is given by
\begin{equation}
\label{visco1}
\eta=\sqrt{\frac{2T }{ M_\pi}} \frac{960 \pi^3 F_\pi^4 }{ 23M_\pi}
g_{7/2}^2\left(z\right){\cal B}_0
\end{equation}
where  ${\cal B}_0$ is a solution of the equation system
\begin{equation}
\label{system} \sum_{s=0}^\infty{\cal B}_s
b_{ts}\left(z\right)=\delta_{t0}
\end{equation}
or
\begin{equation}
\begin{pmatrix}
b_{00}\left(z\right) & b_{01}\left(z\right) & b_{02}\left(z\right) & \cdots \cr
b_{10}\left(z\right) & b_{11}\left(z\right) & b_{12}\left(z\right) & \cdots \cr
b_{20}\left(z\right) & b_{21}\left(z\right) & b_{22}\left(z\right) & \cdots \cr
\vdots & \vdots & \vdots & \ddots \cr
\end{pmatrix}
\begin{pmatrix}
{\cal B}_0 \cr {\cal B}_1 \cr {\cal B}_2 \cr \vdots \cr
\end{pmatrix}
=\begin{pmatrix}
1 \cr 0 \cr 0 \cr \vdots \cr
\end{pmatrix}
\end{equation}
\noindent In order to find the leading contribution to the
viscosity we can truncate the above system to find
\begin{equation}
\label{approx}
 {\cal B}_0=\frac{1}{ b_{00}\left(z\right)}
\end{equation}
In this way the calculation of the viscosity have been reduced to
the computation of the integral $b_{00}\left(z\right)$.

\section{Results and discussion}

According to our previous discussion the viscosity of the pion gas
at low density and temperatures is given by equation
(\ref{visco1}), where ${\cal B}_0$ is the solution of the equation
system given in (\ref{system}) and the corresponding coefficients
are defined in (\ref{bdefinition}). Thus it is possible, at least
in principle, to find the viscosity as a function of the
temperature $T$ and the fugacity $z$ (or the pion chemical
potential). Alternatively it is also possible to find the
viscosity as a function of the temperature and the pion number
density. Due to  the complicated integrals appearing in the
definition of the $b_{st}\left(z\right)$ functions it has been
possible to get numerical results only by Monte Carlo \cite{C}
integration. For the sake of simplicity most of these numerical
results have been obtained by using the approximation of
(\ref{approx}) and the hard sphere cross section of
(\ref{sigma0}). In order to check the validity of these
approximations we have computed the $4\times 4$ matrix containing
the first $b_{st}\left(z\right)$ functions for the particular
value $z=1$. The result and the corresponding errors is
\begin{widetext}
\begin{equation}
b_{st}=\left( \vcenter{ \offinterlineskip\halign{&#\cr $\ 277.2\pm
0.7\ $&\vrule&$\ -42\pm 1\ $&\vrule& $\ -99\pm 3\ $&\vrule&$\ -408\pm
14\ $\cr \hrulefill &\vrule& &\vrule height 3pt && \vrule & \cr &&&
\vrule height 3pt && \vrule & \cr $\ -42\pm 1\ $&&$\ 1043\pm 8\
$&\vrule&$\ -395\pm 8\ $&\vrule&$\ -800\pm 30\ $\cr \hrulefill &&
\hrulefill &\vrule height 3pt && \vrule & \cr &&&&& \vrule height 3pt
& \cr $\ -99\pm 3\ $&&$\ -395\pm 8\ $&&$\ 10700\pm 110\ $&\vrule&$\
-7100\pm 200\ $\cr \hrulefill && \hrulefill && \hrulefill &\vrule
height 3pt & \cr \noalign{\vskip 3pt} $\ -408\pm 14\ $&&$\ -800\pm 30\
$&&$\ -7100\pm 200\ $&&$\ 199000\pm 2000\ $\cr }} \right)
\end{equation}
Now we consider the submatrices $1 \times 1$, $2 \times 2$, $3
\times 3$ and $4 \times 4$ and solve the corresponding system for
the appropriate number of ${\cal B}$ so that we can check the
convergence of the results as the size of the truncation is
increased. The results are shown in table 1. There we see that the
error is of the order of 1.6 \%

\vskip 0.3cm
\centerline{
\vbox{\offinterlineskip \halign{#\hfil&#&&\hfil#\hfil\cr
\noalign{\hrule height 1pt} \noalign{\vskip 1mm} &&${\cal B}_0$&
${\cal B}_1$ &${\cal B}_2$ &${\cal B}_3$\cr \noalign{\vskip 1mm}
&&\hrulefill&\hrulefill&\hrulefill&\hrulefill\cr &\vrule height
6pt&&&&\cr Zero order&\vrule& $3.608\cdot 10^{-3}$&&&\cr &\vrule
height 6pt &&&&\cr First order  &\vrule&$3.630\cdot
10^{-3}$&$1.497\cdot 10^{-4}$& &\cr &\vrule height 6pt &&&&\cr
Second order &\vrule&$3.647\cdot 10^{-3}$&$1.654\cdot
10^{-4}$&$3.985\cdot 10^{-5}$&  \cr &\vrule height 6pt &&&&\cr
Third order &\vrule&$\quad 3.666\cdot 10^{-3}\quad $&$\quad
1.765\cdot 10^{-4}\quad $&$\quad 4.700\cdot 10^{-5}\quad$&$\quad
9.902\cdot 10^{-6}\quad$  \cr &\vrule height 6pt&&&&\cr
\noalign{\hrule height 1pt} }}}
\vskip 0.5cm

To take into account the velocity effects in the cross section we will
consider (\ref{sigma}) instead of (\ref{sigma0}) in the
computation of the $b_{ij}$ integrals. Then we will have
\begin{equation}
b_{st}=b_{st}^0+\frac{T}{ M_\pi}b_{st}^1
\end{equation}
where $b_{ij}^0$ is the leading term corresponding to the hard sphere
approximation and
\begin{eqnarray}
b_{00}^1&=&\int_0^\infty d U d u\int_0^\pi d \theta d \alpha
\int_0^{2\pi} d \phi\ \frac{ \sin \theta \ \sin \alpha \ U^2 u^3
e^{2U^2+\frac{1}{ 2}u^2}}{
\left(e^{{v'_1}^2}-1\right)\left(e^{{v'_2}^2}-1\right)
\left(e^{v_1^2}-1\right)\left(e^{v_2^2}-1\right)} \nonumber \\
&&\left(\frac{13}{ 92}u^2-3U^2+2U^2\cos \theta \frac{2U+u\cos\theta}{
2U\cos\theta+u}\right)
\left(\Delta\left[\left(v_iv_j-\frac{1}{3}\delta_{ij}
v^2\right)\left(1-e^{-v^2}\right)\right]\right)^2
\end{eqnarray}
\end{widetext}
At the zero order in the sense of the previous table we have
\begin{equation}
{\cal B}_0=\frac{1}{ b_{00}}=\frac{1} { b_{00}^0+\frac{T}
{M_\pi}b_{00}^1 }
\end{equation}
which at low temperatures can be written as
\begin{equation}
{\cal B}_0=\frac{1}{ b_{00}^0}\left(1-\frac{T}{ M_\pi}\frac{b_{00}^1}{
b_{00}^0}\right)
\end{equation}
so that
\begin{equation} \eta=\sqrt{\frac{2T }{ M_\pi}} \frac{960 \pi^3 F_\pi^4 }{
23M_\pi} g_{7/2}^2\left(z\right)\frac{1}{
b_{00}^0}\left(1-\frac{T}{ M_\pi}\frac{b_{00}^1}{ b_{00}^0}\right)
\end{equation}
This formula can be applied provided
\begin{equation}
T\ll M_\pi\left|{b_{00}^0\over b_{00}^1}\right|
\end{equation}
After a numerical integration we find, for $z=1$, $b^1_{00}=-310 \pm
40$. By comparison with our previous result $b^0_{00}=277.2\pm 0.7$ we
get the upper bound for the temperature
\begin{equation}
T\ll \left(0.90\pm 0.12\right) M_\pi
\end{equation}
which means that the hard sphere approximation is safe provided
the temperature is much smaller than the $90\%$ of the pion mass.
As we are working  in a non relativistic framework this will
always be the case and then no  additional constrain on the
applicability of our results is coming from the hard sphere
approximation.

Therefore in the following we will work at zero order and in the
hard sphere approximation. In this case the viscosity is given by
\begin{eqnarray}
\eta\left(T,z\right)&=&\sqrt{T}\frac {4096 \sqrt{2} \pi^2 F_\pi^4}{ 345
M_\pi^{3/2}}\nonumber \\ & &\left(b_{00}\left(z\right)\right)^{-1} \left[
\int_0^\infty\frac{x^{5/2}}{ z^{-1}e^x
-1}d x \right]^2
\end{eqnarray}
Some numerical results obtained from this formula are found in table 2
for different values of the fugacity $z$.

\vskip 0.3 cm
\centerline{
\vbox{\offinterlineskip\halign{#\hfil&#&\hfil#\hfil\cr
\noalign{\hrule height 1pt}\noalign{\vskip 1mm} && $\eta
\left(\hbox{MeV}^3\right)$  \cr
\noalign{\vskip 1mm} && \hrulefill \cr
&\vrule height 3pt & \cr
Quantum case ($z=1$) $\quad$ &\vrule&
$\quad \sqrt{T}\ 4.00\cdot 10^5 \quad$ \cr
&\vrule height 6pt & \cr
Quantum case ($z=0.05$) $\quad$& \vrule &
$\quad \sqrt{T} \ 1.40\cdot 10^6\quad$ \cr
&\vrule height 6pt & \cr
Classical case $\quad$ & \vrule&
$\quad \sqrt{T}\ 1.9\cdot 10^6 \quad$ \cr
&\vrule height 3pt & \cr
\noalign{\hrule height 1pt} }}}
\vskip 0.5cm

The classical case is just the limit of $z$ going to zero. This
corresponds to a classical gas of hard spheres of radious
\begin{equation}
 R=\frac{M_\pi}{ \pi F_\pi^2}\sqrt{\frac{23}{ 384}}
\end{equation}

\begin{figure}
\begin{center}
\psfig{figure=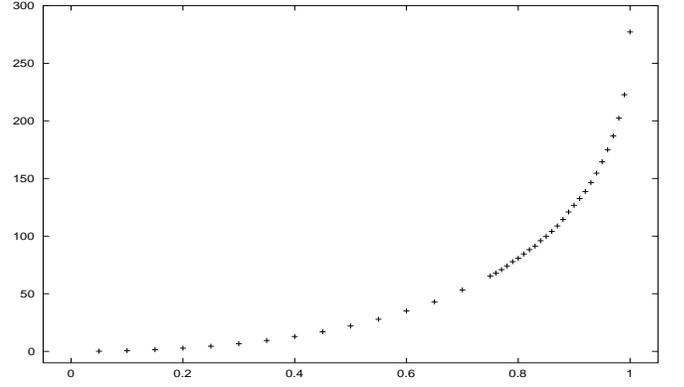,width=8.6cm,height=5cm}
\caption{Plot of
$\left(b_{00}\left(z\right)\right)^{-1}\left[\int_0^\infty\frac{x^{5/2}}{
z^{-1}e^x -1}d x\right]^2$ versus $z$. It can be understood as a plot
of the viscosity, for constant temperature and in arbitrary units,
versus the fugacity.}
\end{center}
\end{figure}

Because of the particular way we have performed our computations
the particular case $z=0$ is not numerically accessible. However
this case has been treated  long time ago and it is possible to
find an analytic result for the viscosity, namely
\begin{equation}
\eta= \sqrt{T}\frac{5\sqrt{M_\pi}}{ 16\sqrt{\pi}R^2}
\end{equation}

Thus we can check that our computations have the proper $z=0$
limit. In fig.1 we show the behaviour of $b_{00}$ and the
viscosity, for constant $T$, in terms of the fugacity. From this
plots we learn that the viscosity of the bosonic quantum gas is
smaller  than that of the classical gas. This could be expected
because of the following heuristic argument: After some elastic
collision the emerging particles have more affinity for occupied
states than classical ones. Therefore the microscopic transfer of
momentum is more effective in a classical gas than in a bosonic
gas and then the viscosity is also larger. A pictorial view of
this fact is displayed in fig.2.

\begin{figure}
\begin{center}
\psfig{figure=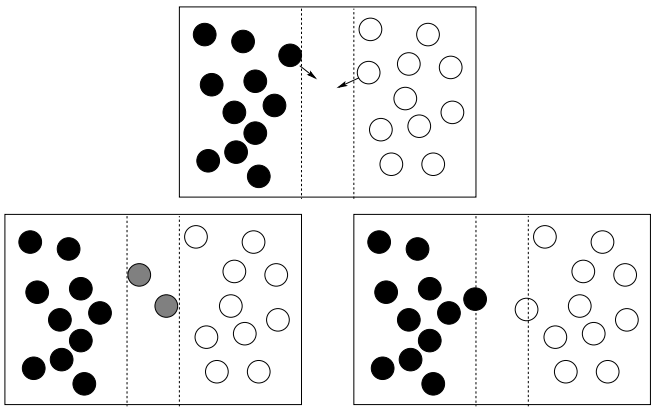,width=8.6cm} \caption{ $a)$ Two particles,
belonging to two different regions of the gas with different
average momentum (black and white), before an elastic collision.
$b)$ Particles before the collision in the classical case where
they typically get a new momentum (represented in grey) $c)$
Particles in the bosonic gas where they typically want to go to
occupied states with the same momentum.}
\end{center}
\end{figure}

In many cases it is useful to have the viscosity as a function of the
temperature and the density instead of the temperature and the
fugacity. The pionic density can be written in terms of the
temperature and the fugacity as
\begin{equation}
\frac { n}{ T^{3/2}} =\frac { g M_\pi^{3/2}}{ \sqrt{2} \pi^2}
\int_0^\infty \frac{x^{1/2}}{ z^{-1} e^x - 1}d x
\end{equation}
This equation defines also implicitly the fugacity as a function
of the density and the temperature $z=z\left(n,T\right)$. However,
due to the complexities of the above integral this functions have
only been computed numerically. In fig.4 we plot $\eta / \sqrt{T}$
as a function of $n/T^{3/2}$. The computed points can be fitted
quite well in the plotted range with a polynomial
\begin{equation}
f_1\left(x\right) = A + B x + C x^2
\end{equation}
with constants $A = 0.1814 \pm 0.0004$, $B = \left(-2.42 \pm
0.03\right)\cdot 10^{-4}$ and $ C = \left(9.6 \pm 0.3\right)\cdot
10^{-8} $. Thus the function obtained (fig.3) for the viscosity
versus the temperature and the density is
\begin{widetext}
\begin{equation}
\label{etavstn}
\eta\left(T,n\right)=\frac{4096 \sqrt{2} \pi^2 F_\pi^4}{ 345
M_\pi^{3/2}}\sqrt{T} \left[0.1814-2.42\cdot 10^{-4}\frac{n}{
T^{3/2}}+9.6\cdot 10^{-8}\left(\frac{n}{T^{3/2}}\right)^2\right]
\end{equation}
\end{widetext}
As it may be noticed in fig.4, the values of the viscosity cover
the range from $10^6$ to $10^7\ \hbox{MeV}^3$ for the considered
densities and temperatures.

\begin{figure}
\begin{center}
\psfig{figure=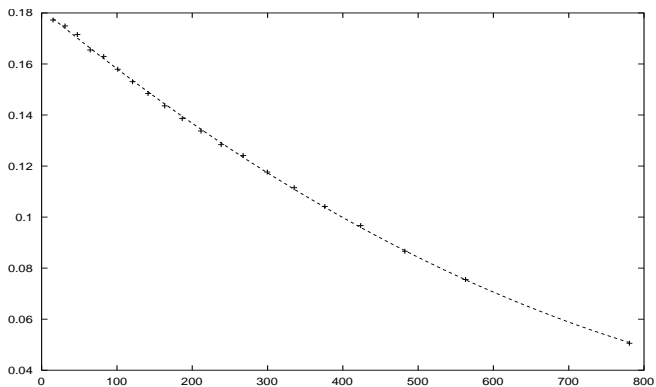,width=8.6cm,height=5cm}
\caption{Numerical fit of $\eta/\sqrt{T}\propto f_1\left(
n/T^{3/2}\right)$.}
\end{center}
\end{figure}

\begin{figure}
\begin{center}
\psfig{figure=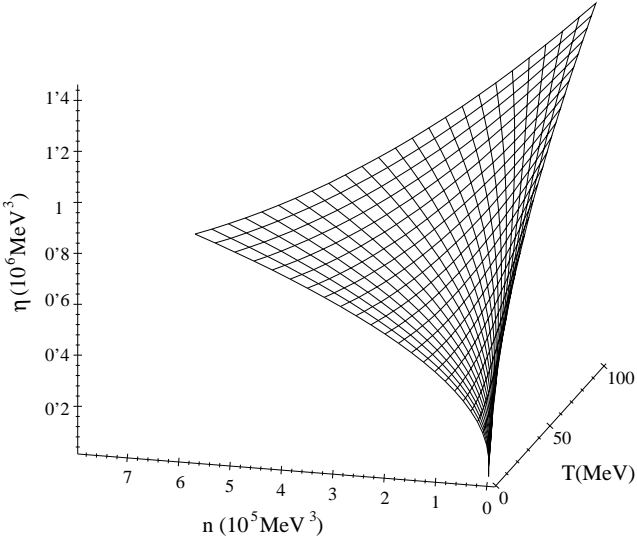,width=8.6cm}
\caption{3D plot of viscosity vs. temperature and particle number
density, given by equation (\ref{etavstn}).}
\end{center}
\end{figure}


\section{Conclusions}

In this work we have computed the viscosity of the pion gas,
starting from first principles only. This computation is relevant
for the hydrodynamic description of hadronic matter at low
energies and densities. The work relies on the use of the
Uehling-Uhlenbeck equation, which is the quantum version of the
Boltzmann equation, and chiral symmetry, which determines
completely the structure of the pion interactions at low energies.
We have also found a formula for the viscosity in terms of the
temperature and  the density which properly fits our numerical
results. The interest of the computations is twofold. First it is
useful to check when the usual assumption of perfect fluid for the
hadronic fluid is reasonable enough. Second it can be used to
include the viscosity in the  Navier-Stokes equations when
dissipative effects cannot be neglected. The main limitation of
our work is that it can be applied only in the non-relativistic
domain. As a consequence of that it cannot be used directly in the
study of the events observed in the modern ultra relativistic
heavy ion colliders such as RHIC or the future LHC. In any case we
understand that our work is interesting in order to show that the
computations of transport coefficients of the pion gas can be done
in a complete model independent way. In fact we consider that the
result presented here is just the first step of a complete
relativistic computation which could be applied at higher energies
and therefore in more realistic situations, to improve the
standard hydrodynamic description of hadronic matter. Work is in
progress in that direction.

\begin{acknowledgments}
The authors want to thank A. Gomez Nicola and J.R. Pelaez for
reading the manuscript. This work has been partially supported by
the Ministerio de Educaci\'on y Ciencia (Spain) (CICYT AEN 97-1693
and PB98-0782).
\end{acknowledgments}

\section{Appendix}

In this appendix we study the main properties of the family of
orthogonal polynomials $P^s_r\left(z;x\right)$ defined on the interval
$\left(0,\infty\right)$ with measure
\begin{equation}
d \mu_r\left(z;x\right)\equiv \frac{x^r}{ z^{-1}e^x -1}
\end{equation}
where  $z\in(0\ 1]$ and $r>0$. By definition the polynomials are
orthogonal for some given $r$, or in other words,
\begin{eqnarray}
\left(P_r^s,P_r^{s'}\right) =\int _0^\infty dx \frac{x^r}{
z^{-1}e^x-1} P_r^s\left(z;x\right) P_r^{s'}\left(z;x\right) \nonumber
\\ = A\left(z;r,s\right)\delta _{ss'}
\end{eqnarray}
For simplicity we define the polynomials so that they are monic,
i.e. the coefficient of the term of highest degree  in each
polynomial is taken to be one. Thus, the first polynomial is
always the unity and his norm is given by
\begin{eqnarray}
P_r^0=1 \rightarrow \left(P_r^0, P_r^0\right)=\int _0^\infty dx \frac{x^r}{
z^{-1}e^x-1}\nonumber \\ = \Gamma\left(r+1\right) g_{r+1}\left(z\right)
\end{eqnarray}
From the condition
\begin{equation}
\left(P_r^0, P_r^1\right)=0
\end{equation}
it is possible to compute the second polynomial which turns out to
be
\begin{equation}
P_r^1\left(z;x\right)=\frac{g_{r+2}\left(z\right)}{ g_{r+1}\left(z\right)}\left(r+1\right)-x
\end{equation}
The third polynomial is obtained from the conditions $\left(P_r^0,
P_r^2\right)=0$ and $\left(P_r^1, P_r^2\right)=0$ which gives
\begin{widetext}
\begin{eqnarray}
P_r^2\left(z;x\right)&=&\frac{\left(r+3\right)g_{r+2}\left(z\right)g_{r+4}
\left(z\right)-\left(r+2\right) g_{r+3}^2\left(z\right) } {
\left(r+2\right)g
_{r+1}\left(z\right)g_{r+3}\left(z\right)-\left(r+1\right)
g_{r+2}^2\left(z\right)}\left(r+2\right)\left(r+1\right)+
\nonumber \\&&
\frac{\left(r+3\right)g_{r+1}\left(z\right)g_{r+4}
\left(z\right)-\left(r+1\right) g_{r+2}\left(z\right)
g_{r+3}\left(z\right)} {\left(r+1\right) g_{r+2}^2\left(z\right)-
\left(r+2\right)g_{r+1}\left(z\right)
g_{r+3}\left(z\right)}\left(r+2\right)x+x^2
\end{eqnarray}
Higher polynomials can be obtained in a similar way.
\end{widetext}

\end{document}